\documentclass[12pt,sort&compress]{article}

\usepackage[dvips]{color}

\usepackage{amsmath}
\usepackage{graphicx}
\usepackage{tabularx}
\usepackage{subcaption}
\usepackage{cite}
\usepackage{amssymb}
\usepackage{mathtools}
\usepackage{lmodern}
\usepackage[sort&compress,numbers]{natbib}
\usepackage[toc,page]{appendix}
\usepackage{multicol}
\usepackage{multirow}
\usepackage[toc,page]{appendix}

\usepackage{tabularx,ragged2e}
\newcolumntype{C}{>{\Centering\arraybackslash}X} 
\usepackage[outercaption]{sidecap}    
\usepackage{hyperref}

\begin{document} 
\title{Local resetting in non-conserving zero-range processes with extensive rates}
\date{}
\author{Pascal Grange\\
{\emph{Division of Natural and Applied Sciences}}\\
{\emph{and Zu Chongzhi Center for Mathematics and Computational Science}}\\
 Duke Kunshan University\\
8 Duke Avenue, Kunshan, 215316 Jiangsu, China\\
\normalsize{{\ttfamily{pascal.grange@dukekunshan.edu.cn}}}}
\maketitle

\begin{abstract}
 A non-conserving zero-range process with extensive 
 creation, annihilation and hopping rates is subjected to local resetting. 
 The model is formulated on a large, fully-connected network of states. The states are equipped with a (bounded) fitness level: particles are added to each state at a rate proportional to the fitness level of the state. Moreover, particles are annihilated at a constant rate, and hop at a fixed rate to a uniformly-drawn state
 in the network. This model has been interpreted in
 terms of population dynamics: the fitness is the reproductive fitness in a haploid population, and the hopping process models mutation.  It has also been interpreted as a model of network growth with a fixed set of nodes (in which particles occupying a state are interpreted as links pointing to this state). In the absence of resetting, the model is known to reach a steady state, which in a certain limit may 
 exhibit a condensate at maximum fitness. If the model is subjected to global resetting by annihilating all particles at Poisson-distributed times, there is no condensation in the steady state. If the system is subjected to local resetting, the occupation numbers of each state are reset to zero at independent random times. These times are distributed according to a Poisson process whose rate (the resetting rate) depends on the fitness. We derive the evolution equation satisfied by the 
  probability law of the occupation numbers. We calculate the average occupation numbers in the steady state. The existence of a condensate 
 is found to depend on the local behavior of the resetting rate at maximum fitness: if the resetting rate vanishes at least linearly at high fitness, a condensate appears at maximum fitness in the limit where the sum of the annihilation and hopping rates is equal to the maximum fitness.

\end{abstract}

\pagebreak

\tableofcontents

\section{Introduction}


 Interacting particle systems with simple microscopic rules on a fully-connected lattice can exhibit rich behavior. In particular, the  
  zero-range process (ZRP), in which particles hop from a given site at a rate that depends only on the occupation number of the site \cite{spitzer5interaction}, gives rise to condensation for a class of decreasing hopping rates worked out in  \cite{godreche2003,condensationDynamics,grosskinsky2003condensation,condensationInhomogeneous,coarsening}.
   In \cite{angel2005critical,nonConserving}, a non-conserving version of the ZRP was introduced, in which 
   particles are removed from each site at a rate increasing (as a power law) with the occupation number of the site, while  particles are added to each site at a 
 constant rate. The phase diagram was found to include a super-extensive high-density phase. \\ 

 In \cite{ZRPSS}, another non-conserving ZRP on a lattice was introduced, in which the particles are allowed to hop to uniformly chosen sites on a fully-connected lattice.  Any particle hops to a uniformly-chosen site in the lattice (at a fixed hopping rate denoted by $\beta$). Moreover, particles die at a uniform rate (this death rate or annihilation rate is denoted by $\delta$), and are created at a rate proportional to the {\emph{fitness level}} of the site they occupy. The motivation was to give a particle version of Kingman's house-of-cards model \cite{KingmanSimple}  of selection and mutation in haploid population dynamics. If the particles are interpreted as bacteria, the fitness is the reproductive fitness (and the hopping rate is the mutation rate, at which bacteria change their fitness to a random value). 
   The non-conserving ZRP introduced in \cite{ZRPSS} can also be mapped to a model of growing  networks with a fixed set of nodes. In this interpretation, the occupation number of a site is the number of links pointing to it from other sites in a large network,  as in the Bianconi--Barab\'asi model \cite{BB,BBCond}. The fitness is the ability of a site to attract new links (and the hopping process corresponds to a random rewiring of the network). The fitness is assumed to be bounded, and its maximum value is set to $1$ (this convention sets the time scale of the model).\\

  The steady state of the model was worked out in \cite{ZRPSS}. If the hopping rate $\beta$ is low enough, the system reaches a steady state, in which the average occupation number $\overline{n_l}(\infty)$ of a fitness level $l$ in $[0,1]$ takes  
 the following form:
\begin{equation}\label{steadyOrdinary}
  \overline{n}_{l}(\infty) =  \frac{l}{\beta + \delta -l} + 
     \frac{ \beta\rho(\infty)q(l)}{\beta + \delta - l},
 \end{equation}
\begin{equation}
  \rho(\infty)  =  \left( 1 - \beta \int_0^1 \frac{q(l) dl}{\delta +\beta -l}\right)^{-1}  \int_0^1 \frac{l \,dl}{\delta + \beta -l},
\end{equation}
\begin{equation}\label{critical}
   {\mathrm{provided}}\;\;\; \;\delta + \beta > 1 \;\;\;\;{\mathrm{and}}\;\;\;\;\;
\beta < \left(  \int_0^1 \frac{q(l) dl}{\delta +\beta -l}\ \right)^{-1}. 
 \end{equation}
 In the above expressions, $q$ is the density of fitness levels (a normalized probability density function on $[0,1]$, vanishing at maximum fitness), and the density $\rho(\infty)=\int_0^1 \overline{n}_{l}(\infty) dl$ is the average occupation number in the steady state. The various  quantities in Eq. (\ref{steadyOrdinary}) will be reviewed in detail in the next section, but we can  observe that $\rho(\infty)$ diverges when $\delta+\beta$ decreases to $1$. This behavior  
  gives rise to an atom at maximum fitness in the distribution of fitness when the sum of the hopping rate and death rate 
  decreases to $1$ (for a fixed hopping rate $\beta< \left(\int_0^1 \frac{q(l) dl}{1-l}\right)^{-1}$):
\begin{equation}\label{densitiesCond}
 \lim_{ \delta \to (1-\beta)^+ }
\frac{\overline{n_l}( \infty )}{\rho(\infty) } = \left(1- \beta \int_0^1 \frac{q(l) dl}{1-l}   \right) \delta_1 + \beta \frac{q(l)}{1- l}.
\end{equation}
 The presence of an atom at high fitness for low mutation rate is a feature  of the house-of-cards model \cite{KingmanSimple} (which is not defined in terms of particles but by a sequence of distributions of fitness in an infinite population).\\

  In \cite{ZRPResetting} the model  was subjected to  resetting by letting it undergo massive-extinction events (all particles are annihilated at Poisson-distributed times).
  Stochastic resetting  to a given initial configuration at Poisson-distributed times has emerged as a powerful tool to induce out-of-equilibrium steady states. Such states  were first computed exactly for models of single-particle systems, including random walkers and active particles \cite{evans2011diffusion,evans2011optimal,pal2015diffusion,kusmierz2014first,evans2018run,refractory,kumar2020active,grange2020susceptibility}(for  reviews, see \cite{topical,gupta2022stochastic}). Interacting particle systems may also be subjected to stochastic resetting.  If all the degrees of freedom in the system are collectively  reset to their initial value, the resetting prescription has been termed {\emph{global resetting}} (for a review of stochastic resetting in interacting particle systems, see \cite{nagar2023stochastic}). Poisson-distributed massive-extinction events considered in  \cite{ZRPResetting} (with a constant rate $r$) correspond to a global resetting. Global resetting has been studied for the Ising model \cite{magoni2020ising}, exclusion processes \cite{basu2019symmetric,sadekar2020zero,karthika2020totally}, as well as predator-prey models \cite{toledo2019predator,evans2022exactly,mercado2018lotka}, fluctuating interfaces \cite{gupta2014fluctuating,gupta2016resetting}, synchronization \cite{sarkar2022synchronization},  reaction-diffusion processes \cite{durang2014statistical}, glassy systems \cite{grange2020entropy}.\\

 The steady-state average occupation number of the fitness level $l$ with resetting rate $r$, (denoted by $\overline{n}_{l,r}(\infty)$ and worked out in \cite{ZRPResetting}) is formally obtained from Eq. (\ref{steadyOrdinary}) by adding the resetting rate to the annihilation rate:
  \begin{equation}\label{resGlobal}
  \begin{split}
  \overline{n}_{l,r}(\infty) =&  \frac{l + r \overline{n_l}(0)}{\beta + \delta +r -l} + 
     \frac{ \beta\rho_r(\infty)q(l)}{ \beta + \delta +r - l},\\
  \rho_r(\infty)  =&    \int_0^1 \frac{l +  r \overline{n_l}(0)}{\beta + \delta +r -l} dl\left( 1 - \beta \int_0^1 \frac{q(l) dl}{\beta + \delta + r -l}\right)^{-1}.  
  \end{split}
 \end{equation}
  At fixed resetting rate, the limit $\beta+\delta \to 1^+$ does not give rise to a 
    condensate, because the fixed rate $r$ as  a regulator.\\

 On the other hand, the degrees of freedom in an interacting particle system can be reset to their initial value independently. Such a resetting  prescription is termed  {\emph{local resetting}}.   Stationary states under local resetting have been explicitly calculated in
  models of binary aggregation \cite{grange2021aggregation}  and exclusion processes  \cite{miron2021diffusion,pelizzola2021simple}, as well as in the voter model \cite{grange2022voter}. In these developments, the rates  of the independent Poisson processes driving the resetting processes were all equal. In this paper we investigate the local resetting of the ZRP with extensive rates, by resetting independently  the occupation number of each state to zero (at Poisson-distributed times).
   Moreover, we allow the resetting rate to depend on the fitness. For instance, the resetting rate could be a decreasing function of the fitness, to attribute a better carrying capacity to states of higher fitness. Allowing such a dependence could lead to a situation where the resetting rate does not act as a regulator, as it does in Eq. (\ref{resGlobal}). It is natural to ask under which condition on the resetting rate an atom can develop at maximum fitness, for a given local-resetting prescription.\\

 The paper is organized as follows.  The  model and the kinetic equations for the occupation numbers of the fitness levels  are reviewed in Section \ref{setNot}. The corresponding  kinetic equations for the occupation numbers of states are worked out in \ref{localResetting}, which allows to define the local-resetting prescription. The induced PDE for the generating function 
 is solved by the method of characteristics in Section \ref{GF}. The steady state is obtained in Section \ref{steadyState}, where the limit $\beta+\delta\to 1$ (already considered in Eq. \ref{densitiesCond}) is studied.  We work out under which conditions on the resetting rate  a condensate may appear at maximum fitness.\\

\section{Review of the model}\label{setNot}

\begin{figure}
\begin{center}
\includegraphics[width=16cm]{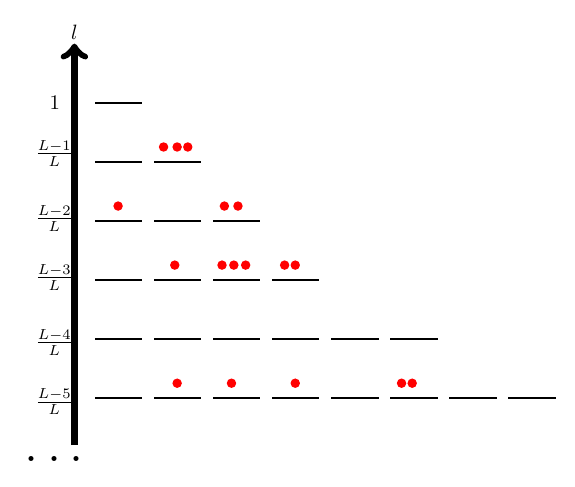} 
\caption{{\bf{A schematic close-up of a configuration showing the occupation of the six highest fitness levels in the system.}} States are shown as horizontal bars, and particles as red dots. Numbers of states shown on the figure are for instance in our notations $v_L = 1$, $v_{L-1} = 2$, $v_{L-3} = 4$, $v_{L-4}= 6$. At fitness level $(L-3)/L$ there are $n_{(L-3)/L} = 6$ particles. The occupation numbers of the corresponding $v_{L-3}$ states are $0,1,3,2$ in this configuration. There is no particle at maximum fitness in this configuration. The rate of appearance of a new particle at the maximum fitness level in this configuration is $1$  according to Eq. (\ref{creationLevel}). There is no particle at fitness level $(L-4)/L$. A particle is added to this fitness level at rate $(L-4)/L$ according to Eq. (\ref{creationLevel}), and sent to one of the $v_{L-4}$ states of fitness $(L-4)/L$ (drawn uniformly). This process is described equivalently 
 in terms of states in Eq. (\ref{creationProcess}): a particle is added to each of the $v_{L-4}=6$ states at rate $1/6$.}
 \label{figConfig}
\end{center}
\end{figure}

The non-conserving zero-range process with extensive rates of \cite{ZRPSS,ZRPResetting} is formulated as a model of particles on a large fully-connected lattice. The vertices  are called {\emph{states}}. Each state can be occupied by any integer number of particles. Moreover, each  state is equipped with a certain level of {\emph{fitness}} (the higher the fitness is, the more particles are produced at the state per unit of time). The fitness levels are in [0,1], and they are regularly spaced by $L^{-1}$, where $L$ is a large integer. Moreover, the number $v_k$ of states at fitness level $k/L$ (for $k$ in $[1..L]$) is defined as:
\begin{equation}\label{vl}
v_k := \mathrm{max}\left( \left[ \frac{1}{L}q\left(\frac{k}{L}\right) V \right], 1\right) = \frac{1}{L}q\left(\frac{k}{L}\right) V + \zeta_k,\;\;\;\;\;\;\;(1\leq k \leq L),
\end{equation}
 where $V$  denotes a large integer, satisfying $V\gg L$, and the square brackets denote the integer part (which implies that the quantity $\zeta_k$ is in the interval $[-1,1]$). The symbol $q$ denotes a fixed probability density, defined on the  continuum $[0,1]$, which is assumed to vanish at high fitness:
\begin{equation}\label{hypoq}
 q(1) = 0, \;\;\;\;\;\;q(l) \underset{l\to 1}{=} O\left( (1-l)^\alpha \right), \;\;\;\;(\alpha>0). 
\end{equation}
 It follows from Eq. (\ref{vl}) that there is exactly one state at the highest fitness level.  A configuration of the system is sketched  on Fig. \ref{figConfig} (showing the highest fitness levels).\\

 The fraction of the total number of states at  fitness level $k/L$ follows from Eq. (\ref{vl}) as:
\begin{equation}
 \frac{v_k}{\sum_{p = 1}^L v_p} = \frac{\frac{1}{L}q\left(\frac{k}{L}\right)  + \frac{\zeta_k}{V} }{ \frac{1}{L}\sum_{p = 1}^L q\left(\frac{p}{L}\right) + \frac{1}{V}\sum_{p=1}^L\zeta_p }.
\end{equation}
 In the large-$L$ limit we recognize a Riemann sum in the denominator
 \begin{equation}
\frac{1}{L}\sum_{p = 0}^L q\left(\frac{k}{L}\right) \simeq \int_0^1 q(l) dl = 1.
 \end{equation}
 Moreover the assumption $V\gg L$ implies
 \begin{equation}
\left| \frac{1}{V}\sum_{p=0}^L\zeta_k \right| \leq \frac{L}{V} = o(1).
 \end{equation}
  In the limit of large $L$, with $V\gg L$, the
  probability density $q$ is therefore the density of states in the interval $[0,1]$ of possible fitness values: 
\begin{equation}\label{densityLargeL}
 \frac{v_k}{\sum_{p = 0}^L v_p} \simeq \frac{1}{L} q\left( \frac{k}{L}\right).
\end{equation}
 
 The dynamics of the model was described in \cite{ZRPSS,ZRPResetting} in terms of the occupation numbers of the fitness levels:\\
 \begin{equation}
 p_l( n,t) := {\mathrm{Prob}}( n\;{\mathrm{particles}}\;{\mathrm{occupy}}\;{\mathrm{fitness}}\;{\mathrm{level}}\;l \;{\mathrm{at}}\;{\mathrm{time}}\; t).
 \end{equation}

 The following three processes contribute to the dynamics:\\
 1. {\bf{Annihilation of particles.}} Particles die at fixed rate $\delta>0$. The total death rate at fitness level $l$ is therefore $\delta n$, where $n$ is the occupation number of this level. The annihilation process therefore induces the following term in the evolution\footnote{Time derivatives are  denoted by  $\frac{d}{dt}$ if time is the only 
 variable taking values in a continuum.} equation of $p_l$:\\
\begin{equation}\label{annihilationLevel}
\left(\frac{dp_l(n,t)}{dt} \right)_{\mathrm{annihilation}} =  - \delta n p_l(n,t) \theta(n)  + \delta (n+1) p_l(n+1,t),
 \;\;\;\;\;\;\;\;(n \in \mathbb{N}).
 \end{equation}
2. {\bf{Creation of particles.}}  Particles are created at a rate proportional to the fitness. The total rate of creation at 
 fitness level $l$ is $l(n+1)$, if $n$ the occupation number of the level, hence the creation term  
\begin{equation}\label{creationLevel}
\left(\frac{dp_l(n,t)}{dt} \right)_{\mathrm{creation}} =  l  n  p_l(n-1,t) \theta(n)  -  l(n+1)  p_l(n,t),\;\;\;\;\;\;\;\;(n \in \mathbb{N}),
\end{equation}
The creation rate is strictly positive for all 
  values of $n$: if $n=0$, a particle is added to fitness level $l$ (this particle occupies a random state, drawn uniformly from the set  of $v_{lL}$ states of fitness $l$). In the network interpretation of the model, this process corresponds to the acquisition of the first link by nodes in a network with a fixed set of nodes (and no existing links to any state of fitness $l$). 
This spontaneous-generation contribution ensures that the configuration with zero occupation number at all states is not a steady state.\\ 
 3. {\bf{Hopping to a uniformly chosen state.}}  Particles hop at a constant rate $\beta>0$, performing a random walk on the network. This process induces\\
\begin{equation}\label{hoppingLevel}
\begin{split}
 \left(\frac{dp_l(n,t)}{dt} \right)_{\mathrm{hopping}} = &\beta \rho(t) q(l) p_l(n-1,t) \theta(n) - \beta  \rho(t) q(l) p_l(n,t) \\
   &   - \beta n p_l(n,t)  \theta(n) + \beta (n+1) p_l(n+1,t),\;\;\;\;\;\;\;\;(n \in \mathbb{N}).\\
\end{split}
 \end{equation}
 where the density $\rho(t)$ is defined as the average occupation number of a fitness level,
\begin{equation}\label{densityDef}
    \rho(t) := \int_0^1 \overline{n_l}(t) dl,  
\end{equation}
 and $\overline{n_l}(t)$ denotes  the average occupation number of  the fitness level $l$:
 \begin{equation}
     \overline{n_l}(t) := \sum_{m\geq 0} m p_l(m,t).
 \end{equation}
 The occurrence of the density $\rho(t)$ in Eq. (\ref{hoppingLevel}) is a consequence of the mean-field geometry. Indeed, the total number of particles hopping per unit of time is $\beta L\rho(t)$. The particles hop to uniformly-drawn states, hence the fitness level $l=k/L$ receives a fraction of the hopping particles equal to the fraction of the states in the system at fitness level.  In the limit of large $L$, and $V\gg L$, this fraction 
 equals the density of states at fitness level $l=k/L$, as explained in Eq. (\ref{densityLargeL}). 
  We therefore obtain the rate at which 
 particles hop to this fitness level:
\begin{equation}\label{fracHopping}
 \beta L\rho(t) \frac{v_k}{\sum_{p = 0}^L v_p}
 \underset{\substack{ L\to \infty,\\ V\gg L}}{\simeq} \beta L \rho(t) \times \frac{1}{L} q\left(l \right) = \beta \rho(t) q\left(l \right),  
\end{equation}
 which reproduces the coefficient in front of the first two terms in  Eq. (\ref{hoppingLevel}).\\

\section{Zero-range process under local resetting}\label{localResetting}

 The evolution  equation of the probability law $p_l$ of the occupation number of fitness level $l$, 
 induced by the processes described in Eqs (\ref{annihilationLevel},\ref{creationLevel},\ref{hoppingLevel}) was solved in \cite{ZRPSS,ZRPResetting}. To subject the model to local resetting, we need to reformulate the  three processes in terms of the 
 occupation number of states. At a fixed fitness level $l=k/L$, the occupation numbers of the $v_k$ states are identically distributed. 
 Let us label the states at this fitness level  by an integer $s$ in $[1..v_k]$. and denote by $\pi_l$ the probability law of the number of particles in a state of fitness $l$:
  \begin{equation}
\begin{split}
 \pi_l(n,t) = {\mathrm{Prob}}( n\;{\mathrm{particles}}\;
     {\mathrm{occupy}}\;{\mathrm{state}}&\;s\;{\mathrm{of}}\;{\mathrm{fitness}}\;l\;{\mathrm{at}}\;{\mathrm{time}}\;t),\\
&\left(l = \frac{k}{L},\;1\leq k \leq L,\; 1\leq s \leq v_k \right).
\end{split}
  \end{equation}

 In the large-$L$ limit, the model is in the mean-field geometry, and  the probability of a list of occupation numbers $(n_1,\dots, n_{v_k})$ of the $v_k$ states at fitness level $l=k/L$ is the product $\prod_{j=1}^{v_k}\pi_l( n_j,t)$.  This factorization property   is a feature of  the dynamics of  urn models studied in   
  \cite{bialas1997,godreche2003,urn1,urn2}.
 Given a total occupation number $M$ of the fitness level, the probability  $p_l( M,t )$ is given as a sum over the partitions of the integer $M$ into $v_k$ nonnegative integers:
\begin{equation}\label{factorization}
p_l( M,t ) = 
\sum_{\substack{\mathbf{n}=(n_1,\dots, n_{v_k} )\in \mathbb{N}^{v_k}\\\sum_{j=1}^{v_k}n_j = M}} 
 \prod_{j=1}^{v_k}\pi_l( n_j,t),\;\;\;\;\;\;\;\;\;\left( M\in \mathbb{N},\;l=\frac{k}{L},\; 1\leq k \leq L \right).
\end{equation}


 Let us  define the 
   creation, annihilation and hopping processes in terms of the quantities $(\pi_{k/L})_{1\leq i \leq v_k}$,
   so that the kinetic equations for the corresponding fitness level (displayed in Eqs (\ref{annihilationLevel},\ref{creationLevel},\ref{hoppingLevel})) are reproduced. \\

\subsection{Annihilation process}
 The particles occupying a given state vanish at a  rate $\delta>0$, hence the annihilation term
\begin{equation}\label{annihilationProcess}
\begin{split}
 \left(\frac{d\pi_l(n,t)}{dt} \right)_{\mathrm{annihilation}} =  - \delta n \pi_l(n,t) \theta(n)  + \delta (n+1) \pi_l(n+1,t),
 \;\;\;\;\;\;\;\left(n \in \mathbb{N}, l = \frac{k}{L},\;1\leq k \leq L\right).
\end{split}
 \end{equation}
 The above equation is intuitively equivalent to Eq. (\ref{annihilationLevel}). The equivalence is derived explicitly in  Appendix \ref{creationRates} using the relation between occupation numbers of states and fitness levels given in Eq. (\ref{factorization}).

\subsection{Creation process}

 The rates of creation of particles are biased by the fitness: let us define the rate of creation
of particles at any state of fitness level $k/L$ 
  to be $(k/L)\times \left( n +  v_k^{-1}\right)$, where $n$ is the occupation number of the state. The constant term $(k/L)v_k^{-1}$ ensures that the process does not stop when all occupation numbers are zero (the factor of $v_k^{-1}$ intuitively corresponds to the $v_k$ choice of states in which to place a particle that is added at rate $k/L$ to the fitness level $k/L$ if it is empty). These rates
  induce the following creation term:\\
\begin{equation}\label{creationProcess}
\begin{split}
\left(\frac{d\pi_l(n,t)}{dt} \right)_{\mathrm{creation}} =  l &\left( n-1 + \frac{1}{v_k} \right)  \pi_l(n-1,k/L,t) \theta(n)  -  l\left( n+\frac{1}{v_k} \right)  \pi_l(n,k/L,t),\\
 &\;\;\;\;\;\;\;\;\;\hspace{5cm}\shoveright{\left(n \in \mathbb{N},\; l = \frac{k}{L},\;1\leq k \leq L\right)}.
\end{split}
 \end{equation}
It can easily be checked using Eq. (\ref{factorization}) that the above equation induces  the creation term displayed in Eq. (\ref{creationLevel}) for the occupation number of the  fitness level $l=k/L$. The corresponding calculations are reported in Appendix \ref{creationRates}.\\

\subsection{Hopping process}

 The particles perform independent random walks on the 
  fully-connected network of states. Each particle hops to a uniformly chosen state, with a fixed rate $\beta$. The fitness level  $k/L$ receives per unit of time a fraction $\beta/L q(k/L) $ of the total number of particles present in the system per unit of time, as explained in Eq. (\ref{fracHopping}).  These particles are shared equally between the $v_k$ states at this fitness level. 
The hopping process induces the following hopping term:
 \begin{equation}\label{hoppingProcess}
\begin{split}
 \left(\frac{d \pi_l(n,t)}{d t} \right)_{\mathrm{hopping}} = &\beta \frac{\rho(t)}{v_k} q(l) \pi_l(n-1,t) \theta(n) - \beta  \frac{\rho(t)}{v_k} q(l) \pi_l(n,t) \\
   &   - \beta n \pi_l(n,t)  \theta(n) + \beta (n+1) \pi_l(n+1,t),\\
   &\;\;\;\;\;\;\;\;\;\hspace{5cm}\left(n \in \mathbb{N},\; l = \frac{k}{L},\;1\leq k \leq L\right).
\end{split}
 \end{equation}
  The above equation induces Eq. (\ref{hoppingLevel}) thanks to Eq. (\ref{factorization}), as explained in Appendix \ref{creationRates}.\\

\subsection{Local-resetting process}

 In the global resetting prescription of \cite{ZRPResetting}, the occupation numbers of all the states in the system are set to zero simultaneously.  These resetting events are Poisson-distributed mass-extinction events. The intensity of the corresponding Poisson process is the constant $r$ that appears in Eq. (\ref{resGlobal}).\\

 In the present model we take a local-resetting prescription, in which the occupation numbers of the states are independently reset to zero. The most general way to pick the intensities of the  Poisson processes generating the resetting times (while ensuring that the occupation numbers at a fixed fitness level are identically distributed), is to make the rate of the Poisson process a function of the fitness. The $v_k$ Poisson processes attached to the states at fitness level $k/L$ are independent, and have the same intensity, which we will denote by $r(k/L)$. 
This local-resetting prescription induces the following terms in the evolution equation of the probability law of the occupation number at a state of given fitness $l$:
\begin{equation}\label{resettingProcess}
\begin{split}
 \left(\frac{d \pi_l(n,t)}{d t} \right)_{\mathrm{resetting}} =& - r(l) \pi_l(n,t) + r(l) \delta_{n,0},\\
&\;\;\;\;\;\;\;\;\;\hspace{3cm}\left(n \in \mathbb{N},\; l = \frac{k}{L},\;1\leq k \leq L\right).
\end{split}
 \end{equation}

The function $r$ is an additional parameter of the model, a nonnegative function defined on $[0,1]$. The parameters of the non-conserving ZRP with extensive rates under local resetting are summarized in Table \ref{tableParams}.
\begin{table}[!ht]
\setlength\extrarowheight{2pt} 
\begin{tabularx}{\textwidth}{|p{1.5cm}|p{3.8cm}|C|C|C|}
   \hline
   {\bf{Symbol}} & {\bf{Values}} & {\bf{Particles}} & {\bf{Network}} & {\bf{Population}} \\
   \hline
   $l$ &   $l\in [0,1]$, {\hbox{approximated}} by $k/L$, with a large integer $L$ and $k$ in $[0..L]$ & rate of {\hbox{production}}  of particles  {{(reproductive fitness)}}     & rate of {\hbox{acquisition}} of links {{(fitness of a node in a network)}} & rate of cell division in a haploid population  {{(reproductive fitness)}}\\
 \hline
 $v_k$ & $v_k \in \mathbb{N}$ & number of states at fitness level $k/L$ & number of nodes at fitness level $k/L$   & number of possible genomes with reproductive fitness $k/L$ \\
\hline
  $L$ &   large integer &total number of distinct  levels       & total number of distinct fitness levels &  total number of distinct fitness levels \\
\hline
   $q(l)$ &   probability density on $[0,1]$, satisfying $q(1)=0$ and $q(l)\underset{l\to 1}{=} O((1-l)^\alpha)$, with $\alpha>0$ & density of states     &  density of states     &   mutant fitness            \\
   \hline
 $\delta$ &  $\delta > 0$ & vanishing rate      & rate of {\hbox{disappearance}} of links             & death rate     \\
\hline
   $\beta$ &  $\beta >0$ & hopping rate &  rewiring rate            & mutation rate         \\
   \hline
  $r(l)$ & nonnegative function of the fitness  & resetting rate of each state of fitness level $l$ to an empty configuration & every node of fitness $l$ loses its links at rate $r(l)$   & the population of every state of fitness $l$ goes extinct at rate $r(l)$\\
 \hline
\end{tabularx}
\caption{Table of definitions and notations for the parameters of the model, with their interpretations in terms of particles, links in a network and haploid population dynamics.}
\label{tableParams}
\end{table}

\subsection{Evolution equation of the occupation number of a state}
Summing the terms described in Eqs (\ref{annihilationProcess},\ref{creationProcess},\ref{hoppingProcess},\ref{resettingProcess}) yields the evolution equation of the probability law of the occupation number of a state of given fitness $k/L$:
 \begin{equation}\label{evolpik}
 \begin{split}
 \frac{d \pi_{k/L}(n,t)}{d t} = \;& \theta( n ) \left\{  \left[ \beta \frac{\rho(t)}{v_k} q\left( k/L \right) +  \frac{k}{L}  \left( n - 1 + \frac{1}{v_k} \right) \right]\pi_{k/L}(n-1,t) 
  - ( \beta + \delta ) n \pi_{k/L}(n,t)    \right\}\\
   &+   (  \beta+ \delta)(n+1)\pi_{k/L}(n+1,t)  -  \left[ \beta \frac{\rho(t)}{v_k} q\left(  k/L\right) +   \frac{k}{L} \left( n+\frac{1}{v_k}\right) + r(k/L)\right]\pi_{k/L}( n,t)\\
   &+ r(k/L)\delta_{n,0}, \;\;\;\;\;\;\;\;\vspace{2cm}(n \in {\mathbb{N}}, 1\leq k \leq L).
\end{split}   
\end{equation}


Let us denote the average number of particles at fitness level $k/L$ by $\overline{n}_{k/L}(t)$. 
  It also depends on the parameters hopping rate $\beta$, annihilation rate $\delta$, and on the density $q$, as well as the resetting rate $r(k/L)$,
  which for brevity is not reflected in the notation.\\
 The average occupation number $\overline{n}_{k/L}(t)$ is the sum of the average occupation numbers at the $v_k$ states of fitness $k/L$:
\begin{equation}
    \overline{n}_{k/L}(t) = \sum_{s=1}^{v_k}\sum_{m\geq 0} m \pi_{k/L}( m, t) = v_k \sum_{m\geq 0} m \pi_{k/L}( m,t). 
\end{equation}

 Let us introduce the generating function of the probability of occupation numbers at a fixed state,
\begin{equation}\label{chiDef}
     \chi(J,k/L,t):= \sum_{n\geq 0 } \pi_{k/L}( n,t)J^n.
 \end{equation}
 The average number of particles at a given fitness level $l=k/L$ 
 (with $v_k=v_{lL}$ states) is therefore expressed in terms of the generating function as
 \begin{equation}\label{nlt}
     \overline{n}_l(t) = v_{lL} \frac{\partial \chi(J,l,t)}{\partial J}|_{J=1}.
 \end{equation}
 The expression of the   density defined in Eq. (\ref{densityDef})  in terms of the generating function follows as
 \begin{equation}\label{rhonl}
\rho(t) =  \int_0^1 dl v_{lL} \frac{\partial \chi(J,l,t)}{\partial J}|_{J=1}.
 \end{equation}
 To obtain the analogue of Eq. (\ref{resGlobal}) in the model with local resetting, we are therefore instructed to  work out and solve the evolution equation of the generating function $\chi$.\\


\section{Generating function}\label{GF}
\subsection{Evolution equation of the generating function}

  Using the identity
 \begin{equation}
 \sum_{n\geq 1} n \pi_{l}( n,t ) J^{n} = J\frac{\partial \chi}{\partial J}( J, l, t),
 \end{equation}
 we obtain the evolution equation of the generating function $\chi$ (defined in Eq. (\ref{chiDef})) as a PDE of order one in $J$ and in time. At a fitness level $l=k/L$, there are $v_k = v_{lL}$ states, and the generating function of the probability law of the occupation number at each of these states satisfies
\begin{equation}\label{timeGenerChi}
\begin{split}
\frac{\partial \chi(J,l,t)}{\partial t}= r( l) +&\left[  (1-J)\left( -\frac{l}{v_{lL}}-\beta \frac{\rho(t)}{v_{lL}}q(l)\right) -r(l)\right]\chi(J,l,t)\\  
&+ (1-J)\left(\beta + \delta - lJ \right) \frac{\partial \chi(J,l,t)}{\partial J}.
\end{split}
\end{equation}
This evolution equation is nonlinear, because 
 the  density $\rho$ is expressed in terms of the generating function (see Eq. (\ref{rhonl})).\\


 As in \cite{ZRPResetting}, we can apply the method of characteristics to solve the PDE, provided the density $\rho(t)$ is treated as a parameter.
  The solution is derived in Appendix \ref{appChar} and reads  
\begin{equation}\label{oriChiBrouillon2}
 \begin{split}
\chi( J, l, t ) =&  \chi\left( 1- \frac{(l-1-\zeta)(1-J)}{l(1-J)-\left( l(1-J) + (1+\zeta - l) \right) e^{(1+\zeta - l )t}} , l, 0\right)\\
 &\times\exp\left( -r(l)t - \frac{(l-1-\zeta)(1-J)}{v_{lL}}\int_0^{t}ds  \frac{  l+\beta\rho(s)q(l)}{l(1-J)-(1+\zeta -lJ)e^{ ( 1+\zeta-l)(t- s)}}   \right)\\
 &+r(l) \int_0^t du \exp\left(  -r(l) u -   
\frac{(l-1-\zeta)(1-J)}{v_{lL}}\int_{0}^{u} dz
  \frac{l+\beta\rho(t-z)q(l)}{l(1-J)-  (1+\zeta -lJ) e^{( 1+\zeta -l)z }}
 \right).
\end{split}
\end{equation}
 where the notation $\zeta$ has been introduced to denote the combination of hopping and annihilation rates:
\begin{equation}
    \zeta := \beta + \delta -1.
\end{equation}
 In these notations (introduced in \cite{ZRPSS}), the annihilation rate is described by the parameter $\zeta$, and the parameter $\delta$ does not appear.
 At this point we have to impose consistency with the definition of the density in Eq. (\ref{rhonl}).

\subsection{Closure condition on the density}

 The average occupation number of the fitness level $l$ at time $t$, expressed Eq. (\ref{nlt}), is readily calculated by Taylor expansion of the generating function:
\begin{equation}
\begin{split}
\chi( J,l,t) = 1 + (J-1)\frac{\partial \chi}{\partial J}(J,l,t)|_{J=1} + o(J-1).
\end{split}
\end{equation}
 The calculation is shown in Appendix \ref{1minusJ} and yields the average occupation number at fitness level $l$ as
 
\begin{equation}\label{nltCLosed}
\begin{split}
 \overline{n_l}(t) = & \overline{n_l}(0)e^{-(1+\zeta+r(l)-l)t}\\
 &+\frac{l}{1+\zeta+r(l)-l}\left( 1-   e^{-(1+\zeta+r(l)-l)t}\right)
+\beta q(l)  \int_0^t \rho( t-s) e^{-(1+\zeta+r(l)-l )s} ds. 
\end{split}
\end{equation}
 The number $v_{lL}$ of states at fitness level $l$ does not appear explicitly in this expression, because each of the $v_{lL}$ states at this fitness level contributes the same average occupation number.\\

Let us denote  the Laplace transform of a function $f$ of time 
 by $\widetilde{f}$:
 \begin{equation}
\widetilde{f}(s) := \int_0^\infty dt e^{-st} f(t). 
 \end{equation}
  The Laplace transform maps convolution products to ordinary products.
 The occupation number 
 of fitness level $l$ is therefore obtained from Eq. (\ref{nltCLosed}) in Laplace space as 
 \begin{equation}\label{densities}
 \begin{split}
 \widetilde{\overline{n_l}}(s) =&  \frac{\overline{n_l}(0)}{1+\zeta+r(l)-l + s}\\
 &+\frac{l}{1+\zeta+r(l)-l}\left( \frac{1}{s}-
 \frac{1}{1+\zeta+r(l)-l +s}\right)
+\widetilde{\rho}(s) \frac{\beta q(l)}{1+\zeta+r(l)-l + s}. 
\end{split}
\end{equation}
Integrating w.r.t. the fitness $l$ and using the definition in Eq. (\ref{densityDef}) yields the closure condition
 on the density in Laplace space
\begin{equation}
\begin{split}
 \widetilde{\rho}(s) =& \int_0^1 dl\frac{ \overline{n_l}(0)}{1+\zeta+r(l)-l + s}\\
 & +\int_0^1 dl\frac{l }{s(1+\zeta+r(l)-l + s)} 
  + \beta \widetilde{\rho}(s) \int_0^1 dl \frac{ q(l)}{1+\zeta+r(l)-l + s}.
 \end{split}
\end{equation}
We therefore obtain the Laplace transform of the density as
\begin{equation}
 \widetilde{\rho}(s) = \frac{1}{s} \frac{s\int_0^1 dl\frac{ \overline{n_l}(0)}{1+\zeta+r(l)-l + s}     +\int_0^1 dl\frac{l }{1+\zeta+r(l)-l + s}}{ 1 - \beta \int_0^1 dl \frac{q(l)}{1+\zeta+r(l)-l + s} }.
\end{equation}
 This expression, which can be substituted into Eq. (\ref{densities}), yields the average occupation 
  of the each fitness level in Laplace space. In particular, it can be used to obtain the steady-state
   values of these occupation numbers.

\section{Steady state of the system}\label{steadyState}
\subsection{Distribution of fitness in the steady state}
If the system reaches a steady state, the density reaches a positive value $\rho(\infty)$ which is given by the final-value theorem:
\begin{equation}
    \rho(\infty) = \underset{s\to 0}{\lim} s \widetilde{\rho}(s).
\end{equation}
  The steady-state value of the density is expressed in terms of the death rate, hopping rate, resetting rate and density of states as 
\begin{equation}\label{rhoInf}
 \rho(\infty) = \frac{1}{1-\frac{\beta}{\beta_c}}\int_0^1 dl\frac{l }{1+\zeta+r(l)-l },
\end{equation}
 which is positive provided $\beta< \beta_c$,
where the  quantity $\beta_c$ is the critical  value of the mutation rate defined as
\begin{equation}\label{betacDef}
\beta_c = \left(    \int_0^1 dl \frac{q(l)}{1+\zeta+r(l)-l} \right)^{-1}.
\end{equation}
 As the function $r$ is nonnegative, all the integrals in Eqs (\ref{rhoInf},\ref{betacDef}) converge if $\zeta>0$. Moreover, the critical value $\beta_c$ is greater than the upper bound on the hopping rate that appears in the  model without resetting (given in Eq. (\ref{critical})):
\begin{equation}
   \left(    \int_0^1 dl \frac{q(l)}{1+\zeta-l} \right)^{-1} \leq   \beta_c.
\end{equation}
 Hence, if the parameters lead to a steady state in the model without resetting (which is the case if $\beta < \left(    \int_0^1 dl \frac{q(l)}{1+\zeta-l} \right)^{-1}$), the condition $\beta < \beta_c$ is satisfied and the system under local resetting reaches a steady state for any choice of the function $r$.\\
 
Applying the final-value theorem to Eq. (\ref{densities})
 yields
 \begin{equation}\label{nlInf}
 \begin{split}
 \overline{n_l}(\infty) =&  \underset{s\to 0}{\lim} s \widetilde{\overline{n_l}}(s) \\
 =&\frac{l}{1+\zeta+r(l)-l}
+ \rho(\infty) \frac{\beta q(l)}{1+\zeta+r(l)-l}.
\end{split}
\end{equation}
 The steady state of the system does not depend on the initial conditions because $\overline{n_l}(0)$ does not appear in Eq. (\ref{rhoInf},\ref{nlInf}).  
The expression of the steady-state average occupation number, average density and critical hopping rate are formally identical to the expression obtained in \cite{ZRPResetting} in the model subjected to global resetting, and reported in Eq. (\ref{resGlobal}).\\


\subsection{Condensation at high fitness in the limit of high density}

 In the ordinary model (without resetting, see Eq. (\ref{densitiesCond})), the density $\rho(\infty)$ goes to infinity and an atom appears at high fitness in the limit where the sum of hopping rate and annihilation rate decreases to the maximum value of the fitness (in our notations this limit is  $\zeta\to 0^+$), at a fixed of the hopping rate satisfying
\begin{equation}
\beta< \left(\int_0^1 \frac{q(l) dl}{1-l}\right)^{-1}.
\end{equation}

 The behavior of the first term in Eq. (\ref{nlInf})  
 in this limit depends on the local behavior of the resetting rate $r(l)$ close to $l=1$.  If the resetting rate vanishes at maximum fitness, this term goes to infinity 
  when $\zeta$ goes to $0$. To identify a limit of high density in the system with local resetting, 
 we must let $r(l)$ go to zero when $l$ goes to the maximum fitness value $1$.  
 Let us assume that the resetting rate vanishes at maximum fitness  with a power-law behavior. There exist two positive constants $C$ and $\nu$ such that
 \begin{equation}
     r(l) \underset{l\to 1}{\sim} C (1-l)^\nu.
 \end{equation}
 Let us distinguish  three cases, depending on whether the resetting rate vanishes linearly ($\nu=1$), sublinearly ($\nu<1$) or superlinearly ($\nu>1$). In all cases, the integral defining the critical value of the hopping rate $\beta$
 defined in Eq. (\ref{betacDef}) has a finite limit when $\zeta$ goes to zero, because of the assumption $q(l)\underset{l\to 1}{=} O((1-l)^\alpha)$ made in Eq. (\ref{hypoq}):\\
\begin{equation}
\underset{\zeta\to 0^+}{\lim} \beta_c^{-1} = \int_0^1 dl \frac{q(l)}{1+r(l) -l}. 
\end{equation}
 Indeed the integrand vanishes at maximum fitness as $(1-l)^{\alpha -1}$ if $\nu\geq 1$, and as  $(1-l)^{\alpha -\nu}\ll(1-l)^{\alpha - 1}$ if $\nu < 1$. Moreover, a fixed hopping rate $\beta< \left( \int_0^1 dl \frac{q(l)}{1+r(l) -l}\right)^{-1}$ satisfies  the condition $\beta< \left( \int_0^1 dl \frac{q(l)}{1+\zeta +r(l) -l}\right)^{-1}$ for $\zeta>0$, which ensures a steady state is reached. Hence we can study the limit $\zeta\to 0^+$ for any such value of $\beta$ using the results obtained in Eqs (\ref{rhoInf},\ref{nlInf}).\\

\subsubsection{Sublinearly vanishing resetting rate at high fitness ($\nu <1$)} 
Consider the case $\nu < 1$. If $\zeta=0$, the denominators in the expression of the average occupation numbers (Eq. (\ref{nlInf})) are equivalent to the resetting rate when the fitness is close to its maximum value:
\begin{equation}
1-l + r(l) \underset{l\to 1}{\sim} C(1-l)^\nu.
\end{equation}
 Hence all the integrals in the expression of the average occupation numbers are finite in the 
 small-$\zeta$ limit:
\begin{equation}
 \begin{split}
 \underset{\zeta\to 0^+}{\lim}\overline{n_l}(\infty)  = & \frac{l}{1+r(l)-l}
+ \rho(\infty) \frac{\beta q(l)}{1+r(l)-l},\\
 \underset{\zeta\to 0^+}{\lim}\rho(\infty) =& \int_0^1 \frac{ldl}{1-l + r(l)} + \int_0^1 \frac{\beta q(l) dl}{1+r(l)-l}.
\end{split}
\end{equation}
The density and all the occupation number have a finite limit when $\zeta$ goes to $0^+$, as in the model with global resetting.

\subsubsection{Linearly vanishing  resetting rate at high fitness}
 The resetting rate has the following behavior
 at high fitness:
 \begin{equation}
r(l) \underset{l\to 1}{\sim} C (1-l).
 \end{equation}

The quantities $1-l$ and $r(l)$ are of the same order close to maximum fitness.
The change of variables from the fitness $l$ to $h$ defined by $l = 1-\zeta h$  yields 
  \begin{equation}\label{stillholds}
 \int_0^1 dl\frac{l }{1+\zeta+r(l)-l } = \zeta\int_0^{\frac{1}{\zeta}}
 \frac{1-\zeta h}{ \zeta h + r(1-\zeta h) + \zeta}.
 \end{equation}

 Maximum fitness corresponds to $h=0$. Taylor expansion close to this value yields
 \begin{equation}
\begin{split}
    \zeta h + r(1-\zeta h) + \zeta =& \zeta\left( 1 + (C+1) h  + o( 1)  \right),\\
    \int_0^1 dl\frac{l }{1+\zeta+r(l)-l } =& \int_0^{\frac{1}{\zeta}} \frac{(1+o(1))dh}{1+(C+1)h}.
\end{split}   
\end{equation}
On the other hand,
\begin{equation}
\begin{split}
\int_0^{\frac{1}{\zeta}} \frac{(1+o(1))dh}{1+(C+1)h} &= \frac{1}{C+1}\int_0^{\frac{1}{(C+1)\zeta}}
 \frac{dz}{1+z}\\ 
&= \frac{1}{C+1}\log\left(   1 + \frac{1}{(C+1)\zeta}\right) =  \frac{1}{C+1}\left[ 
\log(\zeta^{-1} ) + \log\left( \zeta+ \frac{1}{C+1} \right)\right]\\
  &\underset{\zeta \to 0^+}{\sim} -\frac{1}{C+1}\log( \zeta),
\end{split}
\end{equation}
hence the equivalent
 \begin{equation}
\begin{split}
 \int_0^1 dl\frac{l }{1+\zeta+r(l)-l }  \underset{\zeta \to 0^+}{\sim} -\frac{1}{C+1}\log( \zeta).
 \end{split}   
\end{equation}
The density diverges logarithmically when $\zeta$ goes to zero, with a prefactor depending on the local 
 form of the resetting rate at maximum fitness:
 \begin{equation}\label{equivRhoLin}
     \rho(\infty) \underset{\zeta\to 0^+}{\sim} \frac{-\log \zeta}{(C+1)\left(1 - \beta \int_0^1 \frac{q(l) dl}{1-l + r(l)} \right)}.
 \end{equation}
Consider a smooth test function $\phi$ defined on the interval $[0,1]$. Let us study the 
 small-$\zeta$ limit of the integral of $\phi$ against the average occupation number of the steady state:
\begin{equation}
\begin{split}
\int_0^1  \frac{l\phi(l) dl}{1+\zeta+r(l)-l} = &
\zeta\int_0^{\frac{1}{\zeta}}
 \frac{(1-\zeta h)\phi( 1- \zeta h) dh}{ \zeta h + C\zeta h + \zeta}=\phi(1) \int_0^{\frac{1}{\zeta}}\frac{dh}{1 +( C+1)h}\left( 1+o(1)\right)\\
 &\underset{\zeta\to 0^+}{\sim} -\frac{\log(\zeta)}{C+1} \phi(1).\\
 \end{split}
\end{equation}
 \begin{equation}
\begin{split}
\int_0^1 dl \phi(l)  \overline{n_l}(\infty) & \underset{\zeta\to 0^+}{\sim} 
  -\frac{\log(\zeta)}{C+1} \phi(1) 
 -  \beta \frac{\log(\zeta)}{(C+1)\left(1-\beta \int_0^1 \frac{q(l) dl}{1-l + r(l)}\right)} \int_0^1\frac{q(l) \phi(l)dl}{1+r(l)-l}.
\end{split}
\end{equation}
 Hence, dividing both sides by $\rho(\infty)$ and using the equivalent of the density worked obtained in Eq. (\ref{equivRhoLin}) yields
 \begin{equation}
\begin{split}
\frac{1}{\rho(\infty)}\int_0^1 dl \phi(l)  \overline{n_l}(\infty)  \underset{\zeta\to 0^+}{\sim} 
  \left(1-\beta \int_0^1 \frac{q(l) dl}{1-l + r(l)} \right) \phi(1) 
 +\beta \int_0^1\frac{q(l) \phi(l)dl}{1+r(l)-l}.
\end{split}
\end{equation}
 Liberating the test function yields the following limit for the density of occupation numbers (for a fixed hopping rate $\beta< \left(\int_0^1 \frac{q(l) dl}{1+r(l)-l}\right)^{-1}$):
\begin{equation}\label{atomC}
 \frac{1}{\rho(\infty)}  \overline{n_l}(\infty)  \underset{ \delta \to (1-\beta)^+}{\sim} \left(1-\beta \int_0^1 \frac{q(l) dl}{1-l + r(l)} \right) \delta_1
     + \beta \frac{q(l)}{1+r(l)-l}.
\end{equation}
There is a an atom at maximum fitness in the limit of high density.

 \subsubsection{Superlinearly vanishing resetting rate at high fitness ($\nu >1$)}
 
 In this case the resetting rate is subdominant  at high fitness compared to the term $1-l$. 
 For small $\zeta$, let us define again the variable $h$ by $l=1 - \zeta h$. Eq. (\ref{stillholds}) still holds.
 In the limit of small $\zeta$, the  expansion $r(1-\zeta h) = C(\zeta h)^\nu + o(\zeta^\nu)$ induces (at fixed $h$)
\begin{equation}
\begin{split}
    \zeta h + r(1-\zeta h) + \zeta =& \zeta\left( 1 + h + Ch^{\nu} \zeta^{\nu - 1} + o( \zeta^{\nu-1})  \right) = \zeta\left( 1 + h + o(1) \right),\\
    \int_0^1 dl\frac{l }{1+\zeta+r(l)-l } =& \int_0^{\frac{1}{\zeta}} \frac{(1+o(1))dh}{1+h}\underset{\zeta\to 0^+}{\sim} -\log( \zeta).
 \end{split}   
\end{equation}
 On the other hand, the integral in the definition of the critical mutation rate has a finite limit when $\zeta$ goes to zero because the density $q$ goes to zero at high fitness, hence
 \begin{equation}\label{equivRho}
     \rho(\infty) \underset{\zeta\to 0^+}{\sim} \frac{-\log \zeta}{1 - \beta \int_0^1 \frac{q(l) dl}{1-l + r(l)}}.
 \end{equation}
Given a smooth test function $\phi$ defined on the interval $[0,1]$, we obtain:
\begin{equation}\label{eqPhi}
\begin{split}
\int_0^1 dl \phi(l)  \overline{n_l}(\infty) =& 
 \int_0^1  \frac{l\phi(l) dl}{1+\zeta+r(l)-l}
+ \beta\rho(\infty) \int_0^1\frac{q(l) \phi(l)dl}{1+\zeta+r(l)-l}\\
\end{split}
\end{equation}
 Let us apply the same change of variables as above to obtain an equivalent 
  of the first term:
\begin{equation}
\begin{split}
\int_0^1  \frac{l\phi(l) dl}{1+\zeta+r(l)-l} = &
\zeta\int_0^{\frac{1}{\zeta}}
 \frac{(1-\zeta h)\phi( 1- \zeta h) dh}{ \zeta h + r(1-\zeta h) + \zeta}=\phi(1) \int_0^{\frac{1}{\zeta}}\frac{dh}{1 +h}\left( 1+o(1)\right)\\
 &\underset{\zeta\to 0^+}{\sim} -\log(\zeta) \phi(1).\\
 \end{split}
\end{equation}
 Using the equivalent of the average density obtained in Eq. (\ref{equivRho}), we notice that both terms on the r.h.s. of Eq. (\ref{eqPhi}) are of order $-\log \zeta$:
 \begin{equation}
\begin{split}
\int_0^1 dl \phi(l)  \overline{n_l}(\infty) & \underset{\zeta\to 0^+}{\sim} 
  -\log(\zeta) \phi(1) 
 -  \beta \frac{\log(\zeta)}{1-\beta \int_0^1 \frac{q(l) dl}{1-l + r(l)}} \int_0^1\frac{q(l) \phi(l)dl}{1+r(l)-l}.
\end{split}
\end{equation}
 Hence
 \begin{equation}
\begin{split}
\frac{1}{\rho(\infty)}\int_0^1 dl \phi(l)  \overline{n_l}(\infty)  \underset{\zeta\to 0^+}{\sim} 
  \left(1-\beta \int_0^1 \frac{q(l) dl}{1-l + r(l)} \right) \phi(1) 
 +\beta \int_0^1\frac{q(l) \phi(l)dl}{1+r(l)-l}.
\end{split}
\end{equation}
  We read off the distribution of fitness in the steady state in the limit of high density (for a fixed hopping rate $\beta< \left(\int_0^1 \frac{q(l) dl}{1+r(l)-l}\right)^{-1}$):
\begin{equation}\label{resLocal}
    \frac{1}{\rho(\infty)}  \overline{n_l}(\infty)  \underset{ \delta \to (1-\beta)^+}{\sim} \left(1-\beta \int_0^1 \frac{q(l) dl}{1-l + r(l)} \right) \delta_1
     + \beta \frac{q(l)}{1+r(l)-l}.
\end{equation}
 In the limit of high density, the distribution of fitness develops an atom at maximum fitness. The  mass of this atom is expressed  by adding the resetting rate to the combination of hopping and annihilation rates in the expression of the mass of the condensate in the ordinary model, reported in Eq. (\ref{densitiesCond}).

\section{Discussion}

In this work we have worked out in closed form the steady state of a ZRP with extensive rates on a fully-connected lattice, with states subjected to local resetting.  The mean-field geometry allowed to formulate the dynamics in terms of the probability law of the occupation numbers of the states.
 In the local-resetting prescription, the occupation number of any state in the system is set to zero at Poisson-distributed times (the corresponding Poisson process has a rate $r(l)$ that depends only on the fitness level of the state).  In the network interpretation of the model, all links pointing to a given vertex disappear at resetting events. In the population-dynamics interpretation, all bacteria present at a given
  state die at resetting events.\\

The mean occupation numbers of the fitness levels take the same form  as the result obtained in \cite{ZRPResetting}, where the system was subjected to global resetting. The only difference is the dependence of the resetting rate on the fitness level. This formal coincidence emerges in the steady state, but the solution of the evolution equation involves the resetting rate $r(l)$ at every fitness level $l$, whereas the derivation of \cite{ZRPResetting} applied a renewal argument to the solution of the evolution equation in the ordinary model.
  The dependence of the occupation numbers on the resetting rate allows for a condensate to form in the limit where the combined hopping and death rate decrease to the maximum value of the fitness. The formation of an atom depends only on the local behavior of the resetting rate at high fitness: a condensate can form in the limit of high density if the resetting rate $r(l)$ goes to zero at least as fast as $(1-l)$ when the fitness $l$ goes to its maximal value $1$. This result suggests that local resetting may have deep consequences on the phases of an interacting particle system, particularly when the resetting rate depends on the state of the particles. To test this idea in a model with a phase transition, one could for example consider a model of binary aggregation with multiplicative kernel,  for which a gelation transition occurs at finite time \cite{leyvraz1981singularities,hendriks1983coagulation} (see  Chapter 5 of \cite{kineticView} for a review), and reset  polymers  to monomers, at a rate depending on their size.\\

 More realistic versions of the Kingman model include randomness in the mutation rate \cite{Yuan2019,yuan2020kingman}. Alternatively, we could generalize the present model by introducing independent identically-distributed sources of noise to the local resetting rate. The case of unbounded fitness was studied in the limit of large populations in \cite{park2008evolution}. It would be interesting to see whether resetting could interfere with the wave-form of the solution moving towards higher fitnesses.
Local resetting could perhaps be realized experimentally in a modified Lenski experiment \cite{barrick2009genome,Lenski} (in the ongoing Lenski experiment the fitness of a growing bacterial population is studied through regular sampling after exposure to a glucose-based growth medium). Fluctuations in the availability of the nutrient could result in the selective annihilation of bacteria with lower fitness.\\


\begin{appendices}
\section{Evolution equation for the occupation number of a fitness level}\label{creationRates}

 Consider a fixed  fitness level
\begin{equation}
l:= \frac{k}{L},\;\;\;\;\;\;\;\;\;\;\;\hspace{1cm}(k\in[1..L]).
\end{equation}
 The probability law $\pi_l$ of the occupation numbers of each the $v_k$ states at this fitness level is related to the probability law $p_l$ of the occupation number of the entire fitness level $l$ by Eq. (\ref{factorization}). The evolution  equation of  $p_l$ is therefore induced by the evolution equation of $\pi_l$ as follows:
 \begin{equation}
     \frac{dp_l(M,t)}{dt} = \sum_{\substack{\mathbf{n}=(n_1,\dots, n_{v_k} )\in \mathbb{N}^{v_k}\\\sum_{j=1}^{v_k}n_j = M}} \sum_{u=1}^{v_k}\frac{d\pi_l( n_u,t)}{dt} \prod_{\substack{ 1 \leq j \leq v_k  \\ j\neq u}}\pi_l( n_j,t). 
 \end{equation}
 
 Consider the contribution to the r.h.s. of the term proposed in Eq. (\ref{creationProcess}) to model the creation process:
\begin{equation}\label{rato}
\begin{split}
&\sum_{\substack{\mathbf{n}=(n_1,\dots, n_{v_k} )\in \mathbb{N}^{v_k}\\\sum_{j=1}^{v_k}n_j = M}} \sum_{u=1}^{v_k}\left( \frac{d\pi_l( n_u,t)}{dt}\right)_{\mathrm{creation}} \prod_{\substack{ 1 \leq j \leq v_k  \\ j\neq u}}\pi_l( n_j,t)\\
=&
\sum_{\substack{\mathbf{n}=(n_1,\dots, n_{v_k} )\in \mathbb{N}^{v_k}\\\sum_{j=1}^{v_k}n_j = M}} \sum_{u=1}^{v_k} 
\left[ l \left( n_u-1 + \frac{1}{v_k} \right)  \pi_l(n_u-1,l,t) \theta(n_u)  -  l\left( n_u+\frac{1}{v_k} \right)  \pi_l(n_u,l,t)\right]\prod_{\substack{ 1 \leq j \leq v_k  \\ j\neq u}}\pi_l( n_j,t).
\end{split}
\end{equation}

Let us express the following two terms in terms of the
 probability law $p_l$ of the occupation number of the fitness level $l$:

\begin{equation}\label{termn}
\begin{split}
\sum_{ \substack{\mathbf{n}=(n_1,\dots, n_{v_k} )\in \mathbb{N}^{v_k}\\\sum_{j=1}^{v_k}n_j = M} }\sum_{u=1}^{v_k} \Xi_k n_u \pi_k( n_u,t) \prod_{\substack{ 1 \leq j \leq v_k  \\ j\neq u}}\pi_k( n_j,t) =& \sum_{ \substack{\mathbf{n}=(n_1,\dots, n_{v_k} )\in \mathbb{N}^{v_k}\\\sum_{j=1}^{v_k}n_j = M}  }\sum_{u=1}^{v_k} n_u \prod_{1 \leq j \leq v_k}\pi_k( n_j,t)  \\
=&  \Xi_k\sum_{ \substack{\mathbf{n}=(n_1,\dots, n_{v_k} )\in \mathbb{N}^{v_k}\\\sum_{j=1}^{v_k}n_j = M}} \left( \prod_{1 \leq j \leq v_k}\pi_k( n_j,t) \right) 
 \sum_{u=1}^{v_k} n_u\\
=& M  \Xi_k \sum_{\substack{\mathbf{n}=(n_1,\dots, n_{v_k} )\in \mathbb{N}^{v_k}\\\sum_{j=1}^{v_k}n_j = M}} \left( \prod_{1 \leq j \leq v_k}\pi_k( n_j,t) \right) 
 \\
 =& M \Xi_k p_l( M,t), 
\end{split}
\end{equation}

\begin{equation}\label{termConst}
\begin{split}
\sum_{\substack{\mathbf{n}=(n_1,\dots, n_{v_k} )\in \mathbb{N}^{v_k}\\\sum_{j=1}^{v_k}n_j = M}}
\sum_{u=1}^{v_k} \Xi_k \pi_k( n_u,t) \prod_{\substack{ 1 \leq j \leq v_k  \\ j\neq u}}\pi_k( n_j,t) =& \Xi_k \sum_{\substack{\mathbf{n}, \sum_{j=1}^{v_k}n_j = m}}\sum_{u=1}^{v_k}\prod_{1 \leq j \leq v_k}\pi_k( n_j,t)  \\
=& \Xi_k p_l(m,t ), 
\end{split}
\end{equation}
 where $\Xi_k$ is a quantity that may depend on the integer $k$ and on time (like $v_k^{-1}$, $l$, $\delta$ or $\beta \rho(t)$) but not on any of the integers $(n_1,\dots,n_{v_k})$.\\

 Applying the two identities derived in Eqs (\ref{termn},\ref{termConst}), with $l$ and $l/v_k$ substituted to $\Xi_k$, the r.h.s. of Eq. (\ref{rato})
 becomes
\begin{equation}
\begin{split}
  \sum_{\substack{\mathbf{n}=(n_1,\dots, n_{v_k} )\in \mathbb{N}^{v_k}\\\sum_{j=1}^{v_k}n_j = M}} 
\left( \frac{d\pi_l( n_u,t)}{dt}\right)_{\mathrm{creation}} &
\prod_{\substack{ 1 \leq j \leq v_k  \\ j\neq u}}\pi_l( n_j,t)\\
  &=  l\left( M - 1+\frac{1}{v_k}\right) \theta(n) p_l(M-1,t) - l\left( M+\frac{1}{v_k}\right)  p_l(M,t)\\
&= \left(  \frac{dp_l(M,t)}{dt}  \right)_{\mathrm{creation}},
\end{split}
\end{equation}
 which proves that the creation term defined for the fitness level in Eq. (\ref{creationLevel}) is induced by the creation terms for the occupation number of the states as defined in Eq. (\ref{creationProcess}).
 Similarly, substituting the parameters $\delta$, $\beta$ and $\beta \rho(t) v_k^{-1}$ to the quantity $\Xi_k$, we obtain   
\begin{equation}
\begin{split}
  \sum_{\substack{\mathbf{n}=(n_1,\dots, n_{v_k} )\in \mathbb{N}^{v_k}\\\sum_{j=1}^{v_k}n_j = M}} 
\left( \frac{d\pi_l( n_u,t)}{dt}\right)_{\mathrm{annihilation}} 
\prod_{\substack{ 1 \leq j \leq v_k  \\ j\neq u}}\pi_l( n_j,t)
&= \left(  \frac{dp_l(M,t)}{dt}  \right)_{\mathrm{annihilation}},\\
\sum_{\substack{\mathbf{n}=(n_1,\dots, n_{v_k} )\in \mathbb{N}^{v_k}\\\sum_{j=1}^{v_k}n_j = M}} 
\left( \frac{d\pi_l( n_u,t)}{dt}\right)_{\mathrm{hopping}}
\prod_{\substack{ 1 \leq j \leq v_k  \\ j\neq u}}\pi_l( n_j,t)
&= \left(  \frac{dp_l(M,t)}{dt}  \right)_{\mathrm{hopping}}.
\end{split}
\end{equation}
 The expressions proposed in Eqs (\ref{annihilationProcess},\ref{hoppingProcess}) are therefore consistent 
 with the dynamics of the fitness level described by Eqs (\ref{annihilationLevel},\ref{hoppingLevel}).

\section{Characteristic curves for the generating function}\label{appChar}

Let us look for a change of variables from $(J,l,t)$ to some
  new variables $(v,l,t)$, so that the generating function expressed in terms of the new variables satisfies an ordinary differential equation.
  Let us denote by $\hat{\chi}$ the generating function expressed in terms of the new variables:
\begin{equation}\label{functionalDef}
 \hat{\chi}( v, l, t ) := \chi( J(v,l,t), l, t ).
\end{equation}
   
Taking the derivative of Eq. (\ref{functionalDef})  w.r.t. time we obtain:
 \begin{equation}
 \frac{\partial \hat{\chi}( v, l, t )}{\partial t} = \frac{\partial \chi( J(v,l,t), l, t )}{\partial J}  \frac{\partial J(v,l,t)}{\partial t } + \frac{\partial \chi( J(v,l,t), l, t )}{\partial t }.
 \end{equation}
 Inspecting Eq. (\ref{timeGenerChi}), we  impose the following condition so that the coefficient of the derivative $\partial \chi/\partial J$ equals zero:
 \begin{equation}\label{imposed}
\frac{\partial J}{\partial t } = (J-1)\left(\beta + \delta - l J \right).
 \end{equation}
 The average density $\rho$ is a function of time only, and the probability density $q$ is a function of the fitness  only. Their expression does not change when we change variables from $(J,l,t)$ to $(v,l,t)$.  We therefore rewrite
   Eq. (\ref{timeGenerChi})  in the variables $(v,l,t)$ as
  \begin{equation}\label{ODEtrans}
 \frac{\partial \hat{\chi}( v, l, t )}{\partial t} = 
 r(l) -r(l) \hat{\chi}(v, l,t) +\left( 1-J(v,l,t)\right)\left( -\frac{l}{v_k }-\beta \frac{\rho(t)}{v_k}q(l) \right) \hat{\chi}(v, l,t), 
 \end{equation}
 where $l=k/L$, and $v_k$ is the number of states at fitness level $l$.
  The condition obtained in Eq. (\ref{imposed}) is a differential equation, which was solved as follows in \cite{ZRPResetting}. 
   The inverse of the  r.h.s. is a 
  rational function of the parameter $J$, which is easily decomposed into simple elements.
  Assuming that the sum of the creation and annihilation rates is greater than the maximum fitness, we write
  \begin{equation}
    \beta + \delta = 1 + \zeta,
  \end{equation}
 for some $\zeta>0$. The two poles of the rational fraction in Eq. (\ref{ratFrac}) are therefore distinct, and
 \begin{equation}\label{ratFrac}
 \frac{1}{ (J-1)(\beta + \delta -lJ)} = \frac{1}{\beta + \delta-l}\left( \frac{1}{J-1}-\frac{l}{lJ - \beta - \delta} \right).
 \end{equation}
  The quantity $lJ - 1 -\zeta$ is negative if $J$ is in the interval $[0,1]$. Restricting $J$ to this interval,
    we rewrite  Eq. (\ref{imposed}) as follows: 
 \begin{equation}
   \frac{1}{1+\zeta -l}\left( \frac{\partial \log( 1 - J )}{\partial t}  -  \frac{\partial \log( 1 +\zeta -lJ )}{\partial t}  \right)= 1,\;\;\;\;(0\leq J \leq 1),
 \end{equation}
which reads as an equality between time derivatives:
 \begin{equation}
\frac{\partial }{\partial t}\left( \log\left( \frac{1 +\zeta -lJ(v,l,t) }{ 1 - J(v,l,t)}   \right)  \right) = \frac{\partial }{\partial t}\left(   ( l-1-\zeta) t\right).
 \end{equation}
  Integrating w.r.t. time between $0$ and a fixed positive time, denoting by $v$  the integration constant on the r.h.s. yields
  \begin{equation}
\int_0^T dt \frac{\partial }{\partial t}\left( \log\left( \frac{1 +\zeta -lJ(v,l,t) }{ 1 - J(v,l,t)}   \right)  \right) =  \int_0^T dt \frac{\partial }{\partial t}\left(   ( l-1-\zeta) t\right).
 \end{equation}
\begin{equation}\label{vDefConst}
    \log\left( \frac{1 +\zeta -lJ(v,l,t) }{ 1 - J(v,l,t)}  \right)= v + ( l-1-\zeta) t.
 \end{equation}
 The change of variables from $(J,l,t)$ to $(v,l,t)$ is therefore 
 defined by the relation:
   \begin{equation}
    \frac{1 +\zeta -lJ(v,l,t) }{ 1 - J(v,l,t)}   =  e^v e ^{ ( l-1-\zeta) t}.
 \end{equation}
 We can   express the factor of $(1-J)$ needed in Eq. \ref{ODEtrans} in the 
 variables $(v,l,t)$ as:
  \begin{equation}\label{explicitChange}
    1 - J(v,l,t)   =  \frac{l-1-\zeta}{l- e^v e^{ ( l-1-\zeta) t}}.
 \end{equation}

  Substituting into Eq. (\ref{ODEtrans}) yields an explicit form in terms of the variables $(v,l=k/L,t)$ for the time-evolution of the generating function:
 \begin{equation}
 \frac{\partial \hat{\chi}( v, l, t )}{\partial t} =
  r(l) - r(l)\hat{\chi}( v, l, t ) - \frac{l-1-\zeta}{l- e^v e^{ ( l-1-\zeta) t}}
\left( \frac{l}{ v_{lL}}+\beta \frac{\rho(t)}{v_{lL}} q(l)\right)\hat{\chi}( v, l, t ).
 \end{equation}

 Integrating w.r.t. time yields an expression of the generating function as a functional of the density:
\begin{equation}\label{solInv}
 \begin{split}
 \hat{\chi}( v, l, t ) =&  \hat{\chi}( v, l, 0 )\exp\left( -r(l)t - \frac{l-1-\zeta}{v_{lL}}\int_0^{t}ds \left[ l+\beta\rho(s)q(l) \right]  \frac{1}{l-e^v e^{ ( l-1-\zeta) s}}   \right)\\
 &+r(l) \int_0^t ds\exp\left( (s-t) r(l)
  +\frac{l-1-\zeta}{v_{lL}}\int_{t}^s dw
 \left[ l+\beta\rho(w)q(l) \right]  \frac{1}{l-e^v e^{ ( l-1-\zeta) w}}\right).
 \end{split}
 \end{equation}

 To express the generating function in the variables $(J,l,t)$ we have  to express the quantities $e^v$ and $J(v,l,0)$ in terms of $(J,l,t)$.
 Let us rewrite 
 the explicit\footnote{The first equation in the system of Eq. (\ref{solSys}) can also be obtained from the definition of $v$ as an integration constant in Eq. (\ref{vDefConst}), which reads $v = \log\left(  \frac{1+\zeta - J(v,l,0)}{1-J(v,l,0)}  \right)$.} change of variables in Eq. (\ref{explicitChange}), at time  $0$ and at time $t$:
 \begin{equation}\label{solSys}
 \begin{cases}
    1 - J(v,l,0)   =&  \frac{l-1-\zeta}{l- e^v},\\
    1- J(v,l,t) =& \frac{l-1-\zeta}{l- e^v e^{(l-1-\zeta)t}}.
\end{cases}    
 \end{equation}
 Solving the above system in the unknowns $e^v$ and $J(v,l,0)$, we obtain 
 an expression of $e^v$ in terms of the quantities $J(v,l,t)$, $l$ and $t$,
 from which we find
\begin{equation}
\begin{split}
   e^v =& \frac{-lJ(v,l,t) + 1 + \zeta}{1-J(v,l,t)}e^{-(l-1-\zeta)t},\\
   J(v,l,0)   =& 1- \frac{l-1-\zeta}{l-\left( l + \frac{1+\zeta - l}{1-J(v,l,t)} \right) e^{(1+\zeta - l )t} }.
\end{split}   
\end{equation}
Moreover,  the  functional relation in Eq. (\ref{functionalDef}) implies
\begin{equation}\label{apro}
  \hat{\chi}( v, l, 0 ) = \chi( J(v,l,0), l, 0 ) =
  \chi\left( 1- \frac{l-1-\zeta}{l-\left( l + \frac{1+\zeta - l}{1-J(v,l,t)} \right) e^{(1+\zeta - l )t}} , l, 0\right).
\end{equation}
Substituting Eqs (\ref{solSys},,\ref{apro}) into Eqs (\ref{solInv}), 
we obtain the desired expression of the generating function in the variables $(J,l,t)$:
\begin{equation}\label{solInvEnd}
 \begin{split}
 \chi( J, l, t ) =&  \chi\left( 1- \frac{l-1-\zeta}{l-\left( l + \frac{1+\zeta - l}{1-J(v,l,t)} \right) e^{(1+\zeta - l )t}} , l, 0\right)\\
 &\times\exp\left( -r(l)t - \frac{l-1-\zeta}{v_{lL}}\int_0^{t}ds  \frac{  l+\beta\rho(s)q(l)}{l-\frac{1+\zeta -lJ}{1-J}e^{ ( 1+\zeta-l)(t- s)}}   \right)\\
 &+r(l) \int_0^t ds\exp\left( (s-t) r(l)
  +\frac{l-1-\zeta}{v_{lL}}\int_{t}^s dw
  \frac{l+\beta\rho(w)q(l)}{l-  \frac{1+\zeta -lJ}{1-J} e^{( 1+\zeta-l)(t- w) }}\right).
 \end{split}
 \end{equation}
 Changing integration variables in the last term from $(s,w)$ to $(u,z)$ defined by $t-w = z$ and $-s+t=u$ yields
\begin{equation}
\begin{split}
\int_0^t ds\exp&\left( (s-t) r(l)
  +\frac{l-1-\zeta}{v_{lL}}\int_{t}^s dw
  \frac{l+\beta\rho(w)q(l)}{l-  \frac{1+\zeta -lJ}{1-J} e^{( 1+\zeta-l)(t- w) }}\right)\\
&= \int_0^t du \exp\left(  -r(l) u -   
\frac{l-1-\zeta}{v_{lL}}\int_{0}^{u} dz
  \frac{l+\beta\rho(t-z)q(l)}{l-  \frac{1+\zeta -lJ}{1-J} e^{(1+\zeta -l)z }}
 \right).
\end{split}
 \end{equation}
Rearranging yields
\begin{equation}\label{solInvEnd}
 \begin{split}
 \chi( J, l, t ) =&  \chi\left( 1- \frac{(l-1-\zeta)(1-J)}{l(1-J)-\left( l(1-J) + (1+\zeta - l) \right) e^{(1+\zeta - l )t}} , l, 0\right)\\
 &\times\exp\left( -r(l)t - \frac{(l-1-\zeta)(1-J)}{v_{lL}}\int_0^{t}ds  \frac{  l+\beta\rho(s)q(l)}{l(1-J)-(1+\zeta -lJ)e^{ ( 1+\zeta-l)(t- s)}}   \right)\\
 &+r(l) \int_0^t du \exp\left(  -r(l) u -   
\frac{(l-1-\zeta)(1-J)}{v_{lL}}\int_{0}^{u} dz
  \frac{l+\beta\rho(t-z)q(l)}{l(1-J)-  (1+\zeta -lJ) e^{( 1+\zeta -l)z }}
 \right).
 \end{split}
 \end{equation}

This is the result reported in Eq. (\ref{oriChiBrouillon2}).

\section{Average occupation number of a state}\label{1minusJ}

Using Eq. (\ref{oriChiBrouillon2}), we can express $\chi( 1+h,l,t)$ as follows:
\begin{equation}\label{oriChiBrouillonRep}
 \begin{split}
 \chi( 1+h, l, t ) =& \chi\left(1+ \frac{h (1+\zeta - l)}{hl+\left( -h l + (1+\zeta - l) \right) e^{(1+\zeta - l )t}}, l, 0 \right)\\
 &\times \exp\left( -r(l)t + \frac{ h(1+\zeta -l)}{v_{lL}}\int_0^{t} ds \frac{ l + \beta\rho(s)q(l) }{hl + (1+\zeta - l -lh) e^{ ( l-1-\zeta) (s-t)}}    \right)\\
 &+r(l) \int_0^t du \exp\left( -r(l) u
  +\frac{h( l- 1 -\zeta )}{v_{lL}}\int_{0}^{u}
dw \frac{ l +\beta \rho(t - w) q(l) }{ -hl - (1+\zeta -l -lh) e^{ ( 1+\zeta-l) w}}\right).
\end{split}
\end{equation}
In particular, using the normalization condition of $\pi_l$ at time $t=0$, which reads $\chi\left(1, l, 0 \right)=1$ we can check that
\begin{equation}\label{oriChiBrouillonRepoo}
 \begin{split}
 \chi( 1, l, t ) =& \chi\left(1, l, 0 \right)\times \exp\left( -r(l)t  \right)+r(l) \int_0^t du \exp\left( -r(l) u \right)\\
 =& \exp\left( -r(l)t  \right) + 1 - \exp\left( -r(l)t  \right) = 1,
\end{split}
\end{equation}
 which is the normalization condition of $\pi_l$  at time $t$.\\

The Taylor expansion of Eq. (\ref{oriChiBrouillonRep}) at order one in $h$ of $\chi( 1+h, l, t )$ therefore reads
\begin{equation}\label{oriChiBrouillonRepo}
 \begin{split}
 \chi( 1+h, l, t ) - 1=&  +he^{-(1+\zeta - l) t}\frac{\partial \chi( 1,l,0)}{\partial J}e^{-r(l)t}\\
&+\frac{h}{v_{lL}}e^{-r(l) t} \int_0^t ds [l+\beta\rho(s) q(l)]e^{(1+\zeta - l)(s-t)}\\
 &+\frac{h r(l)}{v_{lL}}  \int_0^t du e^{-r(l)u} \int_0^{u} dw  [l+\beta\rho(t-w) q(l)]
  e^{ -( 1+\zeta -l) w} + o(h)\\
  =&  +he^{-(1+\zeta - l) t}\frac{\partial \chi( 1,l,0)}{\partial J}e^{-r(l)t}\\
&+\frac{l h}{v_{lL}}e^{-r(l) t} \int_0^t ds e^{(1+\zeta - l)(s-t)} + \frac{l r(l) h}{v_{lL}}  \int_0^t du e^{-r(l)u} \int_0^{u} dw  
  e^{ -( 1+\zeta -l) w}\\
 &+\frac{h \beta q(l)}{v_{lL}}e^{-r(l) t} \int_0^t ds \rho(s) e^{(1+\zeta - l)(s-t)}
  +\frac{h r(l) \beta q(l)}{v_{lL}} \int_0^t du e^{-r(l)u} \int_0^{u} dw  \rho(t-w)
  e^{ -( 1+\zeta -l) w} \\
   &+ o(h).
\end{split}
\end{equation}
Let us permute the integrals in the last double integral containing the density:
\begin{equation}\label{t1}
\begin{split}
r(l) \int_0^t du e^{-r(l)u} \int_0^{u} dw  \rho(t-w)
  e^{ -( 1+\zeta -l) w} =& \int_0^{t} dw \rho(t - w) e^{ -( 1+\zeta -l) w} \left(   r(l) \int_{w}^t du e^{-r(l)u} \right)\\
  =& \int_0^{t} dw \rho(t -w) e^{-(1+\zeta - l)w} \left( e^{-r(l)w} - e^{-r(l)t} \right)\\
  &= (\rho\ast W_{1+\zeta + r(l) - l}) - e^{-r(l) t} \left( \rho \ast W_{1+\zeta-l}\right)(t),
  \end{split}
\end{equation}
 with the notation $W_a(t):= e^{-at}$. On the other hand,
  defining a new integration variable $p$ by $s=t-p$ (at fixed $t$)
\begin{equation}\label{t2}
\int_0^t ds \rho(s) e^{(1+\zeta - l)(s-t)}
= \int_0^t dp \rho(t-p) e^{-(1+\zeta-l) p}=\left( \rho \ast W_{1+\zeta-l}\right)(t).
\end{equation}
 The remaining integrals in Eq. (\ref{oriChiBrouillonRepo}), which do not involve the density $\rho$, are readily calculated taking the same steps as above:
 \begin{equation}\label{t3}
 \begin{split}
e^{-r(l)t}\int_0^t ds e^{(1+\zeta - l)(s-t)} +& r(l)\int_0^t du e^{-r(l)u} \int_0^{u} dw e^{-(1+\zeta-l)w}\\
=& e^{-r(l)t}\int_{p}^0 dw e^{-(1+\zeta - l)p} + \int_{0}^t dw e^{-(1+\zeta - l)w} \left( -e^{-r(l) t} +e^{r(l)w}  \right)\\
 =& \int_{0}^t dw e^{-(1+\zeta +r(l) - l)w}\\
 =& \frac{1}{1+\zeta + r(l) - l}\left[ 1-  e^{-(1+\zeta +r(l) - l)t}\right].
 \end{split}    
 \end{equation}
 
 Substituting Eqs (\ref{t1},\ref{t2},\ref{t3}) into Eq. (\ref{oriChiBrouillonRepo}) yields
the average occupation number of a state at fitness level $l$ and time $t$: 
\begin{equation}
\begin{split}
\frac{\partial\chi( J, l, t )}{\partial t}|_{J=1} =& +e^{-(1+\zeta +r(l)- l) t}\frac{\partial\chi( J, l, 0 )}{\partial t}|_{J=1}\\
&+
\frac{1}{v_{lL}}\frac{l}{1+\zeta + r(l) - l}\left[ 1-  e^{-(1+\zeta +r(l) - l)t}\right]+ \frac{\beta q(l)}{ v_{lL}}
( \rho\ast W_{1+\zeta + r(l) -l})(t).
\end{split}
\end{equation}
 Multiplying by the number of states $v_{lL}$ at fitness level $l$
  yields the average occupation number at 
   fitness level $l$ reported in Eq. (\ref{nltCLosed}).


\end{appendices}

\bibliography{bibRefsNewJun2023} 
\bibliographystyle{ieeetr}

\end{document}